# Seasonal association between viral causes of hospitalised acute lower respiratory infections and meteorological factors in China: a retrospective study

*Bing Xu\*, Jinfeng Wang†, Zhongjie Li\*, Chengdong Xu, Yilan Liao, Maogui Hu, Jing Yang, Shengjie Lai, Liping Wang, Weizhong Yang†*

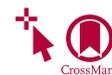

## Summary

**Background** Acute lower respiratory infections (ALRIs) caused by respiratory viruses are common and persistent infectious diseases worldwide and in China, which have pronounced seasonal patterns. Meteorological factors have important roles in the seasonality of some major viruses, especially respiratory syncytial virus (RSV) and influenza virus. Our aim was to identify the dominant meteorological factors and to model their effects on common respiratory viruses in different regions of China.

**Methods** We analysed monthly virus data on patients hospitalised with ALRI from 81 sentinel hospitals in 22 provinces in mainland China from Jan 1, 2009, to Sept 30, 2013. We considered seven common respiratory viruses: RSV, influenza virus, human parainfluenza virus, adenovirus, human metapneumovirus, human bocavirus, and human coronavirus. Meteorological data of the same period were used to analyse relationships between virus seasonality and seven meteorological factors according to region (southern vs northern China). The geographical detector method was used to quantify the explanatory power of each meteorological factor, individually and interacting in pairs, on the respiratory viruses.

**Findings** 28 369 hospitalised patients with ALRI were tested, 10 387 (36·6%) of whom were positive for at least one virus, including RSV (4091 [32·0%] patients), influenza virus (2665 [20·8%]), human parainfluenza virus (2185 [17·1%]), adenovirus (1478 [11·6%]), human bocavirus (1120 [8·8%]), human coronavirus (637 [5·0%]), and human metapneumovirus (615 [4·8%]). RSV and influenza virus had annual peaks in the north and biannual peaks in the south. Human parainfluenza virus and human bocavirus had higher positive rates in the spring–summer months. Human metapneumovirus had an annual peak in winter–spring, especially in the north. Adenovirus and human coronavirus exhibited no clear annual seasonality. Temperature, atmospheric pressure, vapour pressure, and rainfall had most explanatory power on most respiratory viruses in each region. Relative humidity was only dominant in the north, but had no significant explanatory power for most viruses in the south. Hours of sunlight had significant explanatory power for RSV and influenza virus in the north, and for most viruses in the south. Wind speed was the only factor with significant explanatory power for human coronavirus in the south. For all viruses, interactions between any two of the paired factors resulted in enhanced explanatory power, either bivariately or non-linearly.

**Interpretation** Spatiotemporal heterogeneity was detected for most viruses in this study, and interactions between pairs of meteorological factors were found to enhance their influence on virus variation. These findings might be helpful to guide government planning, such as public health interventions, infection control practice, and timing of passive immunoprophylaxis, and might facilitate the development of future vaccine strategies.

**Funding** National Natural Science Foundation of China, the Ministry of Science and Technology of China, and the Technology Major Project of China.

## Introduction

Acute lower respiratory infections (ALRIs) are a major public health problem in both high-income and lower-income countries, causing nearly 2·38 million deaths globally in 2016,[1] making them the fifth leading cause of death overall and the leading infectious cause of death in all-age deaths over the past three decades.[2–4] A large majority of ALRI fatalities occur in low-income and middle-income countries, with 40% reported in Africa and 30% in southeast Asia.[5]

Many pathogens are responsible for ALRIs including bacteria, viruses, and fungi. Most respiratory viruses responsible for ALRIs were detected for the first time in recent years,[6–9] such as respiratory syncytial virus (RSV), influenza viruses, human parainfluenza virus, adenovirus, human metapneumovirus, human rhinovirus, human coronavirus, and human bocavirus. Among these viruses, RSV is the most predominant viral cause of hospital admissions, especially in infants, being responsible for almost 65% of hospitalised cases.[10] As the

**Research in context**

**Evidence before this study**
We searched Web of Science, Google Scholar, and China National Knowledge Infrastructure on Sept 30, 2019, with no date or language restrictions, using the following broad search terms: "ALRI", "respiratory", "influenza OR respiratory syncytial virus OR RSV OR parainfluenza OR PIV OR metapneumovirus OR MPV OR adenovirus OR ADV OR coronavirus OR hCoV OR human bocavirus OR hBoV". Our search yielded more than 200 studies; more than 50 of these studies were related to the association between meteorological factors and acute lower respiratory infections (ALRIs) or common respiratory viruses, with most focusing on respiratory syncytial virus (RSV) and influenza virus in some local regions. Some evidence suggested that epidemics of RSV and influenza virus were associated with cold environmental temperature and relative humidity in temperate regions and with rainfall in tropical regions. However, most previous studies have lacked a systematic analysis to objectively compare a range of meteorological factors including other respiratory viruses (such as human parainfluenza virus, adenovirus, human metapneumovirus, human bocavirus, and human coronavirus). There is, therefore, a need for studies using long-term data on a larger spatial scale.

**Added value of this study**
To our knowledge, this is the first systematic analysis of the quantitative contributions of different meteorological factors and their interactions to the seasonality of the seven most common respiratory viruses in both temperate (north) and subtropical (south) climate regions in China. The seasonality of RSV, influenza virus, and human metapneumovirus is clear, with peaks in winter in the north and some summer peaks in the south, whereas adenovirus, human parainfluenza virus, human bocavirus, and human coronavirus do not exhibit clear annual seasonality. We find that temperature, atmospheric pressure, vapour pressure, and rainfall have strong associations with ALRI hospital admissions in both northern and southern China; relative humidity had significant explanatory power for some viruses in the north but not in the south. By contrast, hours of sunlight influenced circulation only for RSV and influenza viruses in the north, but for the majority of viruses in the south. Importantly, the single contributions of these factors are difficult to extract because their interaction powers are non-linearly or bivariately enhanced.

**Implications of all the available evidence**
Our findings can be used to predict the seasonal patterns of these common respiratory viruses and interpret the mechanism of how meteorological factors act on them. More understanding of the meteorological effects on the activity of these viruses would also be helpful in guiding government planning, such as public health interventions, infection control practice, and timing of passive immunoprophylaxis, and might facilitate the development of future vaccine strategies.

second most common virus, influenza virus can cause widespread morbidity and mortality among human populations worldwide. The infection mechanism of parainfluenza virus is similar to that of RSV infection in children, and serological surveys have indicated that nearly every child is likely to be infected at least once with RSV or human parainfluenza virus before age 5 years.[11]

The seasonality of ALRIs has been found to vary across different study periods and regions in different countries; recently, a global seasonal pattern of some common viruses has been described in detail.[12] A clear understanding of the relationship between seasonality of different viruses and meteorological factors is key to the successful implementation of prevention and control programmes. In recent years, many studies have focused on how weather patterns influence the seasonality and transmission of RSV and influenza virus in some local regions.[13–17] However, most previous studies have lacked a comprehensive evaluation of weather factors on other common respiratory viruses simultaneously.

China is a geographically and climatologically diverse country with a temperate climate in the north, subtropical climate in the south, and a tropical climate in Hainan and some cities in Guangdong. In this study, using data on the viral causes of hospitalised cases of ALRI covering 5 consecutive years in most representative regions in China, we provide insight into the influence of different meteorological factors on common respiratory viruses in different major climate regions. If the explanatory powers of these weather factors can be determined, the seasonality and transmission of these viruses in different environments can be better understood, which would help in the prediction of future outbreaks and government planning.

## Methods

### Viral assessment of hospitalised cases of ALRI
In this study, active surveillance for all-age hospitalised patients with ALRI in 108 sentinel hospitals in 24 provinces of China was initiated from Jan 1, 2009, to Sept 30, 2013. Included hospitals were chosen after carefully considering capacities of surveillance and laboratory testing, and for geographical representativeness. Specimens from the first two or five ALRI patients admitted weekly, or monthly, in each sentinel hospital were tested for RSV, influenza A virus, human parainfluenza virus (types 1–4), adenovirus, human coronavirus (OC43, 229E, and NL63), human bocavirus, and human metapneumovirus. Verbal consent was recorded from patients or their parents or guardians during their enrolment. Detailed data regarding patient demographics and clinical characteristics including signs and symptoms and laboratory test results were collected. Virus detection methods included nucleic acid testing, virus isolation, and serological testing. Cases of ALRI were defined by a national surveillance protocol developed by the Chinese Centre for Disease Control and Prevention and regional reference laboratories.[18] After excluding 27 sentinel hospitals that had very few ALRI cases (<40 in total), we enrolled 28369 hospitalised patients with ALRI from 81 sentinel hospitals (in 36 cities) in 22 provinces in China for the final analysis (figure 1). To align with the current surveillance and research situation in China, we divided China into northern and southern parts by conventional geographical divisions, following the Qinling Mountain range to the west and the Huai River to the east.[19]

## Outcomes

The monthly positive rates of seven common viruses in patients with ALRI were calculated for the study period to represent the seasonal activity of each virus in each region as follows:

$$V_{ij} = \frac{\text{number of positive tests}_{ij}}{\text{number of patients hospitalised with ALRI}_{ij}}$$

where $i=1, 2,...,12$ (from January to December) and $j=1, 2,...,5$ (from 2009 to 2013), and $V$ represents each of the seven virus types (influenza virus, RSV, human parainfluenza virus, adenovirus, human metapneumovirus, human coronavirus, and human bocavirus). Details on the seven respiratory viruses are shown in table 1, along with their potential climate influence factors, which can act on the seasonal characteristics of the viruses through various mechanisms (figure 2).

## Meteorological data

24 daily meteorological parameters were obtained from the China Meteorological Data Sharing Service System for 756 ground weather stations nationwide from Jan 1, 2009, to Sept 30, 2013. Considering the multicollinearity in some meteorological parameters—eg, maximum, minimum, and mean atmospheric pressure have a strong correlation of 0·99—only seven average daily parameters were included in this analysis: mean temperature, mean atmospheric pressure, vapour pressure, rainfall, hours of sunlight, mean relative humidity, and mean wind speed. The results of other meteorological parameters were omitted. Monthly meteorological data were calculated by averaging the daily value of each meteorological variable in each month, except for rainfall and hours of sunlight, which were calculated by summation. City-level monthly meteorological indicators for the study area were interpolated by the inverse distance weighted interpolation method. Monthly meteorological data for the whole of China were obtained by averaging data for the 36 study cities, and the monthly meteorological data for northern and southern

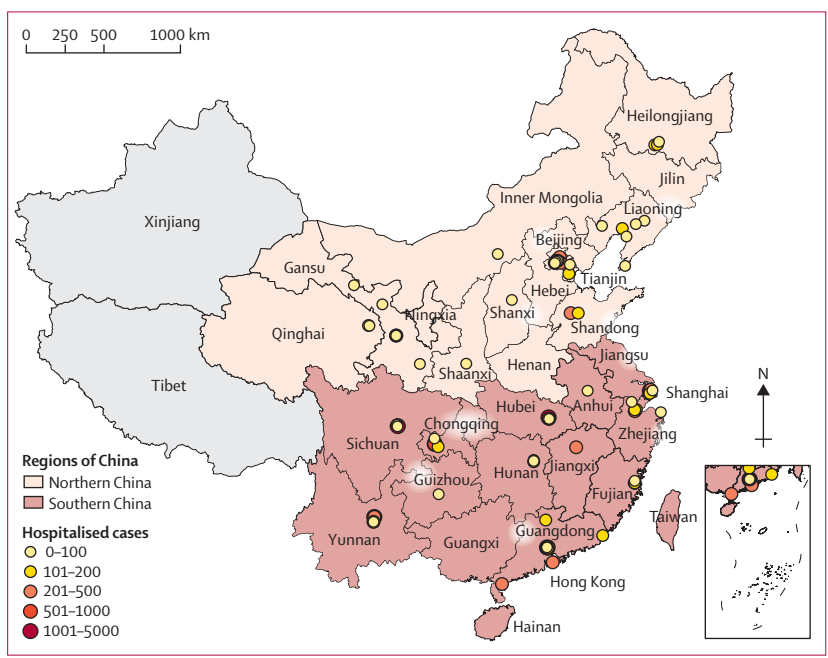

*Figure 1:* Location of 81 surveillance hospitals for hospitalised cases of ALRI
Xinjiang and Tibet are excluded here. Circle size and colour represent the number of cases contributed by each hospital, with bigger and darker circles indicating more cases. ALRI=acute lower respiratory infection.

China were calculated by averaging those cities in the north or the south, respectively.

## Statistical analysis

We first did a descriptive analysis by plotting the time-series of the rates for each respiratory virus and each meteorological factor in both northern and southern China. We used Student's $t$ test to compare variables between regions (eg, number of tests or monthly rates). The effects of individual meteorological factors and their interactive effects were evaluated by the geographical detector method. The geographical detector, proposed by Wang and colleagues,[27,28] is a spatial variance analysis method originally designed as a test of spatial stratified heterogeneity and is now widely used to explore the explanatory power of driving factors responsible for heterogeneity. Spatial stratified heterogeneity is a widely existing phenomenon that describes that within-strata variance is less than between-strata variance classification.[29] If an occurrence, such as ALRI incidence, is dominated by certain risk factors, the occurrence and the risk factors will share a similar spatial and temporal distribution, which is the key underlying assumption used in the geographical detector. Unlike traditional linear models, the geographical detector is not restricted by the assumption of linearity and immunity to the collinearity multivariable. Here, we use the $q$-statistic in the geographical detector to express the influence of each meteorological factor on virus variation.[27] The $q$-statistic represents the explanatory power of a factor $X$ on an outcome $Y$, which can be interpreted as $X$ explaining

For the **China Meteorological Data Sharing Service System** see http://data.cma.gov.cn

| | Family | Structure | Climate influences |
|---|---|---|---|
| Respiratory syncytial virus[13,14,16,20,21] | Paramyxoviridae | Non-segmented, negative sense, single-stranded RNA virus | Low temperature, relative humidity |
| Influenza virus[13,15,17,22] | Orthomyxoviruses | Lipid-enveloped, single-stranded, pleomorphic RNA virus | Low temperature, low relative humidity |
| Human parainfluenza virus[20,23,24] | Paramyxoviridae | Enveloped, single-stranded negative sense RNA virus | Temperature, relative humidity, wind speed |
| Adenovirus[23,25] | Mastadenovirus genus in the family Adenoviridae | Non-enveloped, double-stranded, icosahedral DNA virus | Temperature |
| Human bocavirus[20] | Members of the genus Bocavirus of the family Parvoviridae, subfamily Parvovirinae | Non-enveloped, single-stranded DNA virus | Low temperature, wind speed |
| Human coronavirus[20,23] | Subfamily Coronavirinae in the family Coronaviridae and the order Nidovirales | Enveloped, positive-sense, single-stranded RNA virus | Temperature |
| Human metapneumovirus[20,26] | Belongs to the order Mononegavirales, the genus Metapneumovirus, the subfamily Pneumovirinae | Enveloped, non-segmented negative-strand RNA viruses or mononegaviruses | Low temperature, wind speed |

*Table 1:* Summary of respiratory viruses and potential climate influence factors from the literature

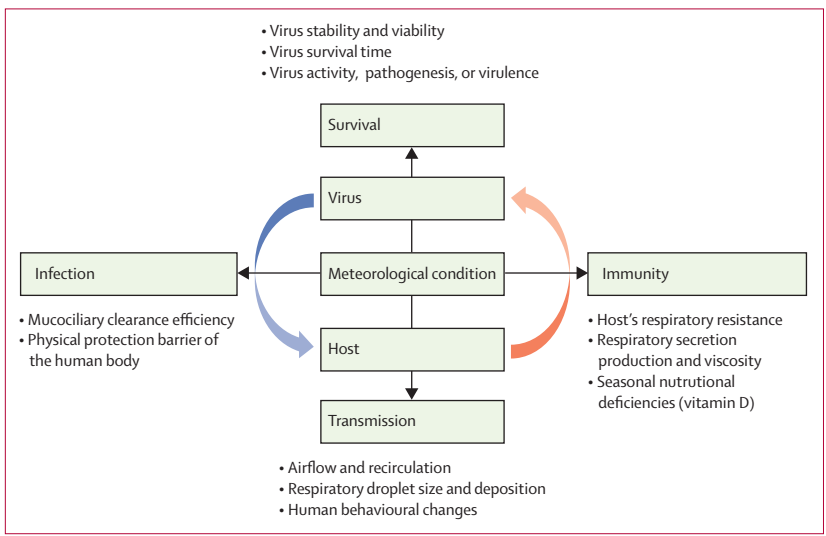

*Figure 2:* Mechanism of the actions of meteorological factors on ALRI viruses
ALRI=acute lower respiratory infection.

$100q\%$ of $Y$. The mathematical formulation of the $q$-statistic is as follows:[29,30]

$$q = 1 - \frac{SSW}{SST},$$

where $SST = N\sigma^2$ and $SSW = \sum_{h=1}^{L} N_h \sigma_h^2$

$q \in [0,1]$

For the **geographical detector packages and software** see http://www.geodetector.cn

See **Online** for appendix 2

where $N$ represents the total number of samples (in this case, monthly positive rates of each virus) and $\sigma^2$ the variance of $Y$ (the monthly positive rate of each virus in each region) in the whole study area; $h = 1, 2, ..., L$ is the number of strata (groupings of months) of the meteorological factors $X$; and $N_h$ is the mean number of samples and $\sigma_h^2$ the local variance of $Y$ in strata $h$. Therefore, SSW is the within sum of squares and SST the total sum of squares. The range of the $q$-statistic is from 0 to 1; the larger $q$ is, the stronger the influence of factor $X$ is on $Y$. If $q$ approaches 1, the value of $\sigma_h^2$ is close to 0, meaning that $X$ has the same distribution as $Y$. Note that $q$-statistics of different factors are not additive, as they each represent the factor's explanatory power on $Y$ separately.

In addition, an interaction detector can be used to reveal the interactive effects of every pair of different meteorological factors $(X_1, X_2)$. Let $q(X_1)$ and $q(X_2)$ be the $q$-statistics of factors $X_1$ and $X_2$, respectively. By overlaying $X_1$ and $X_2$, we can calculate the $q$-statistic of their interaction, $q(X_1 \cap X_2)$, and by comparing this with $q(X_1)$ and $q(X_2)$, we can assess whether their interaction weakens or enhances each factor's effect on ALRIs, or whether the two factors are independent. If $q(X_1 \cap X_2) > q(X_1)$, then $X_2$ enhances $X_1$ (and vice versa if $q(X_1 \cap X_2) > q(X_2)$); if $q(X_1 \cap X_2)$ is greater than both $q(X_1)$ and $q(X_2)$, the factors mutually (bivariately) enhance each other; and if $q(X_1 \cap X_2) > q(X_1) + q(X_2)$, the factors enhance each other non-linearly. If $q(X_1 \cap X_2) < q(X_1)$, then $X_2$ weakens $X_1$ (and vice versa if $q(X_1 \cap X_2) < q(X_2)$); if $q(X_1 \cap X_2)$ is smaller than both $q(X_1)$ and $q(X_2)$, the factors are non-linearly weakened by one another; and if $q(X_1 \cap X_2) < q(X_1) + q(X_2)$, the factors mutually weaken each other. Finally, if $q(X_1 \cap X_2) = q(X_1) + q(X_2)$, then the factors are independent of each other.

All statistical analyses were done using R (version 3.6.0). A two-sided p value of less than 0·05 was considered statistically significant. The geographical detector method was implemented using R packages and software available online.

### Role of the funding source
The funder had no role in study design, data collection, data analysis, data interpretation, or writing of the report.

### Results
Demographic and clinical characteristics of the patients included in this study are shown in appendix 2 (p 2). Of the 28 369 patients tested, 17 127 (60·4%) were younger than 5 years of age. There were more male patients with ALRI than female patients (male-to-female ratio 1·79:1

|  | Temperature | Atmospheric pressure | Vapour pressure | Rainfall | Hours of sunlight | Relative humidity | Wind speed |
|---|---|---|---|---|---|---|---|
| Overall | 0·552* | 0·439* | 0·601* | 0·531* | 0·525* | 0·373† | 0·216 |
| 0–4 years | 0·395 | 0·328† | 0·460 | 0·433 | 0·431 | 0·352 | 0·154 |
| 5–64 years | 0·498† | 0·489* | 0·554† | 0·540† | 0·422 | 0·374 | 0·148 |
| ≥65 years | 0·208 | 0·233 | 0·192 | 0·162 | 0·193 | 0·132 | 0·146 |
| North | 0·632* | 0·478* | 0·639* | 0·615* | 0·268† | 0·415* | 0·151 |
| 0–4 years | 0·503* | 0·434* | 0·557* | 0·407 | 0·398* | 0·395† | 0·175 |
| 5–64 years | 0·428† | 0·453* | 0·401 | 0·493† | 0·364 | 0·378 | 0·149 |
| ≥65 years | 0·274 | 0·317† | 0·367† | 0·282† | 0·407 | 0·330† | 0·279 |
| South | 0·476* | 0·453* | 0·495* | 0·382* | 0·325† | 0·092 | 0·108 |
| 0–4 years | 0·323 | 0·287 | 0·326 | 0·353 | 0·330 | 0·333 | 0·312 |
| 5–64 years | 0·481† | 0·325 | 0·555* | 0·417† | 0·404† | 0·412* | 0·176 |
| ≥65 years | 0·326 | 0·174 | 0·129 | 0·261 | 0·202 | 0·109 | 0·056 |

Data are q-statistics. *p<0·01. †p<0·05.

*Table 2*: Explanatory power of meteorological factors on total respiratory viruses, by age group and region of China

overall), which remained the case for patients with an identified viral infection (ratio 2·02:1) and for each virus type (appendix 2 p 2). 10 387 (36·6%) patients tested positive for at least one respiratory virus (monoinfection and co-detection). The most frequently detected virus was RSV, which was identified in 4091 (32·0%) positive samples, followed by influenza virus (2665 [20·8%]), human parainfluenza virus (2185 [17·1%]), and adenovirus (1478 [11·6%]), with human bocavirus (1120 [8·8%]), human coronavirus (637 [5·0%]), and human metapneumovirus (615 [4·8%]) occupying a small percentage. 22 218 (78·3%) hospitalised patients with ALRI were in southern China (appendix 2 p 3), and the absolute number of virus-positive tests in northern and southern China was significantly different (p=0·016), as was the monthly positive rates of the viruses in the two regions (p=0·004; appendix 2 p 4). Statistical descriptions of the included respiratory viruses and meteorological variables, by region, are shown in appendix 2 (p 3).

Monthly positive rates of each virus and meteorological factors by region are shown in appendix 2 (pp 4–5). The average monthly positive rate for any virus was 45% in China, peaking annually between November and the following March; of the years considered, only 2012 showed a decline in test-positive rates in December instead of an increase, whereby the onset of the epidemic period was delayed by 2 months (appendix 2 p 4). When considering meteorological factors, seasonality in the south was similar to China as a whole, while in the north, the difference between the epidemic period and non-epidemic period was more pronounced (appendix 2 p 5). Almost all viruses could be detected year round, and the three most common viruses were RSV, influenza virus, and human parainfluenza virus (appendix 2 p 4).

There was a clear seasonal variation in RSV-associated hospital admissions, with peaks in the winter months of November–March and very low positive rates during the summer months of June–September (appendix 2 p 4). In the summers of 2009, 2011, and 2013, there was a small peak in the south, and the onset of the epidemic period was delayed by 2 months in 2010 in both the north and south. Influenza had annual peak in cold months in winter in the north and biannual peaks in summer and winter in the south (appendix 2 p 4). By contrast, human parainfluenza virus and human bocavirus rates were higher in spring–summer in most research years (appendix 2 p 4). Adenovirus and human coronavirus exhibited no clear annual seasonality. Although seasonality was not obvious for human metapneumovirus, an annual peak was observed in winter–spring (ie, January–May), especially in the north (appendix 2 p 4).

When considering the explanatory power of meteorological factors on total virus-positive rates in different age groups, we found that almost all meteorological factors, except for hours of sunlight, had greater influence on ARTI admissions in the north than the south (table 2). Vapour pressure was the most dominant factor in each region. The results differed by age group and sex (appendix 2 p 6). Overall, the explanatory power of each meteorological factor was greater in males than in females for each age group. In people aged 65 years or older, significant associations between meteorological factors and any virus were only observed for northern China.

In the whole of China, all included meteorological factors had significant influence on RSV and human metapneumovirus (figure 3). For RSV, temperature had the greatest explanatory power ($q=0·654$), followed by vapour pressure ($q=0·622$) and atmospheric pressure ($q=0·575$). Vapour pressure ($q=0·424$) was the most dominant factor for human metapneumovirus. Temperature, atmospheric pressure, vapour pressure, rainfall,

Articles

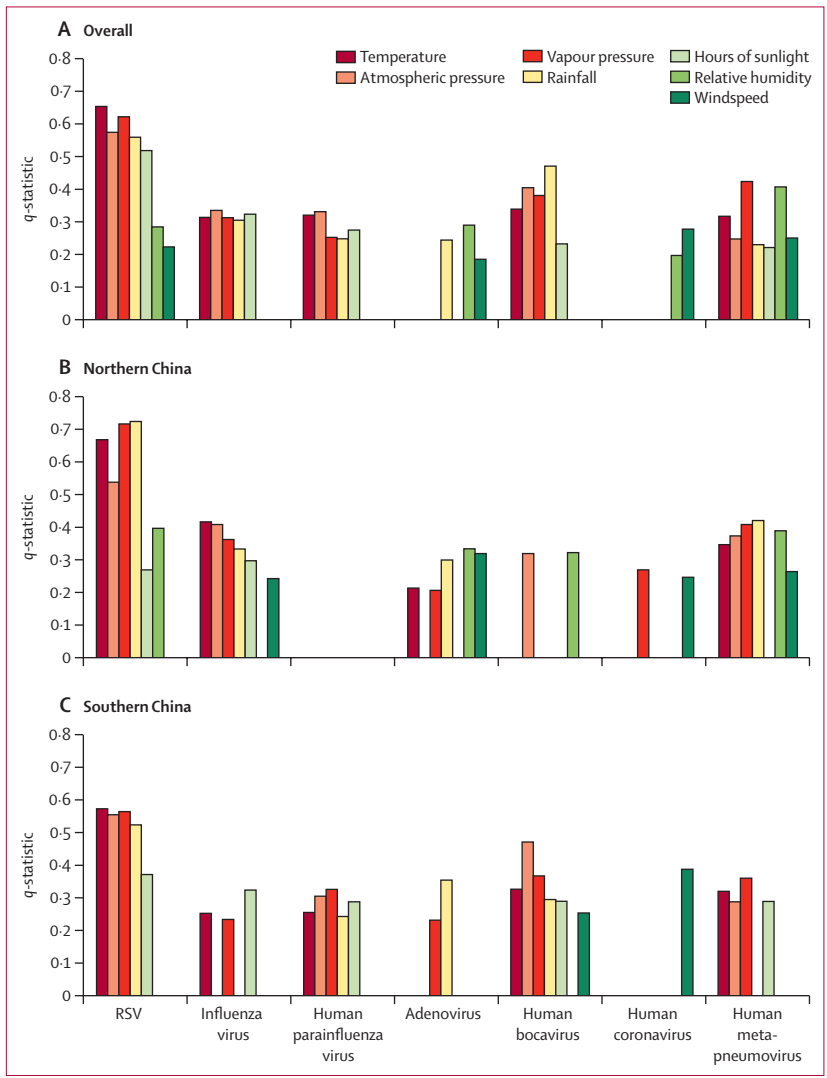

*Figure 3:* Explanatory power of meteorological factors on each respiratory virus in different regions of China
Only meteorological factors with significant explanatory power (p<0.05) are shown. RSV=respiratory syncytial virus.

and hours of sunlight had significant explanatory power for influenza virus, human parainfluenza virus, and human bocavirus, whereas wind speed and relative humidity were significantly associated with adenovirus and human coronavirus. Rainfall was also a dominant power for adenovirus.

In northern China, all factors except for wind speed significantly influenced the variation of RSV, with rainfall the dominant factor ($q=0.723$; figure 3). Temperature was the dominant factor for influenza virus ($q=0.416$); the remaining factors could all individually explain more than 25% of influenza virus variation except for relative humidity. No significant explanation was found for human parainfluenza virus. Relative humidity was the dominant factor for adenovirus ($q=0.333$) and human bocavirus ($q=0.322$). For human metapneumovirus, all factors except for hours of sunlight had high explanatory power. However, human coronavirus was associated only with vapour pressure and wind speed.

In southern China, relative humidity had no significant influence on any of the viruses (figure 3). RSV, human parainfluenza virus, and human bocavirus were significantly associated with temperature, atmospheric pressure, vapour pressure, rainfall, and hours of sunlight. Temperature, vapour pressure, and hours of sunlight had significant influences on influenza virus. Adenovirus was significantly associated with vapour pressure and rainfall. Wind speed was the only factor with explanatory power for human coronavirus, being able to explain 39% of the variation. The dominant factor for human metapneumovirus was vapour pressure ($q=0.358$). Results by age group and sex are shown in appendix 2 (pp 7–11), alongside the variation trend of each and total virus positive rates on the dominant meteorological factor in each region (appendix 2 pp 12–19).

Figure 4 shows the interactive effects of each paired meteorological factors on total respiratory viruses in different regions (see appendix 2 pp 20–26 for each virus). The dominant interaction (ie, largest $q$-statistic) in different regions is shown in appendix 2 (p 27). Overall, the explanatory power of any two independent factors was enhanced after interaction, either bivariately or non-linearly (appendix 2 pp 20–26). When considering the whole of China, the dominant interaction differed by virus. For example, the dominant interactive effect for RSV was wind speed and atmospheric pressure ($q=0.836$), which was greater than the sum of each individual effect ($q=0.224$ for wind speed and $q=0.575$ for atmospheric pressure), while the dominant interactive effect for influenza virus was vapour pressure and hours of sunlight ($q=0.697$, with $q=0.313$ for vapour pressure and $q=0.324$ for hours of sunlight). In the north, relative humidity played a great role in the interaction effect (figure 4). For most viruses, the dominant interactive power was relative humidity interacting with other factors, except for adenovirus, whose dominant interactive effect was rainfall and wind speed (appendix 2 p 23). Although the independent explanatory power of wind speed was not significant for most viruses, its interactive effect with other factors was sometimes the dominant one, especially in the south (figure 4; appendix 2 p 27).

## Discussion

To our knowledge, this study is the first to quantitatively show the influence of different meteorological factors and their interactions on seven common respiratory viruses causing ALRIs, using a large sample of cases across regions of temperate and subtropical climate in China. This study highlights how the seasonality of common respiratory viruses varies, and the different explanatory powers of various meteorological variables on these respiratory viruses across major climate regions of China. We found that influences of meteorological factors could partly explain seasonal variations of most of these viruses.

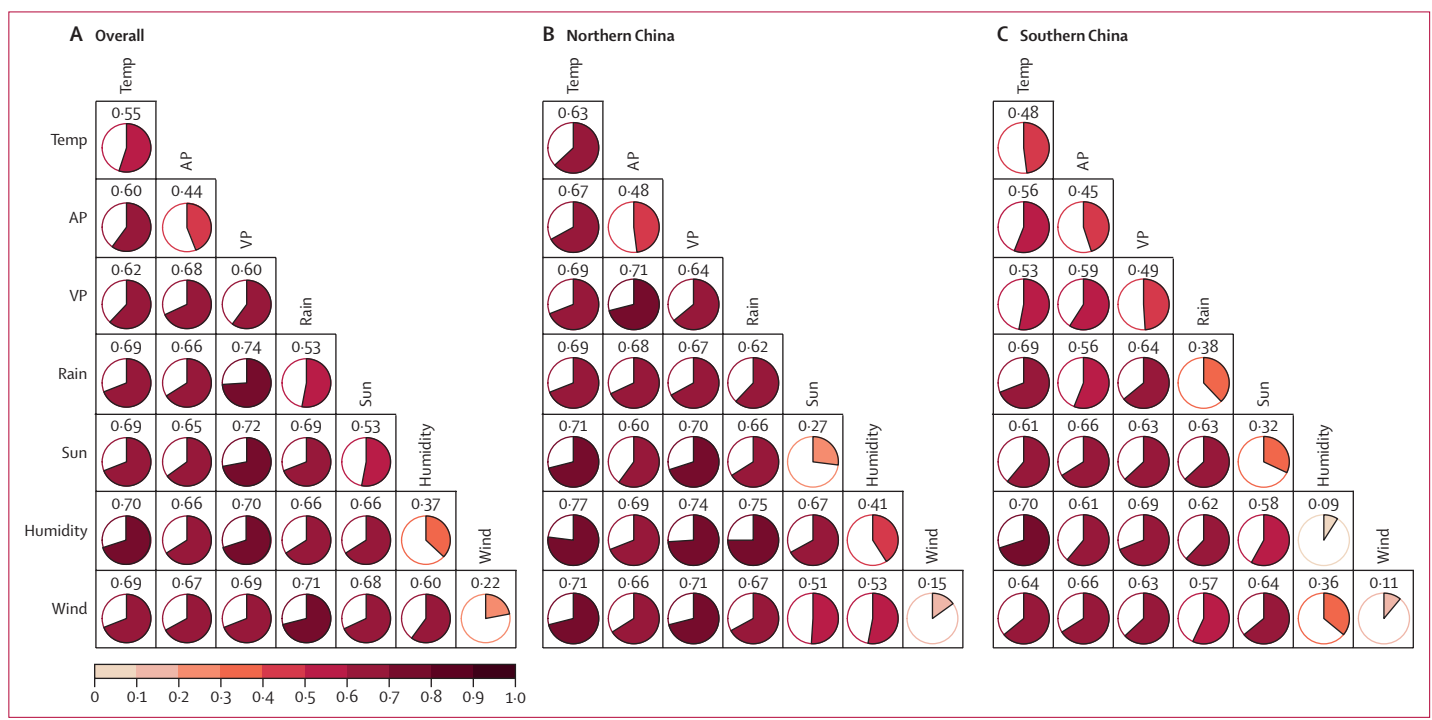

*Figure 4:* Interactive effects of paired meteorological factors on total virus-positive rates, by region
Values on the diagonal show *q*-statistics for the individual factors without interactions. Temp=temperature. AP=atmospheric pressure. VP=vapour pressure, Rain=rainfall. Sun=hours of sunlight. Humidity=relative humidity. Wind=wind speed.

Overall, our findings on epidemic cycles of respiratory viruses in different regions of China were consistent with previous studies in other temperate and subtropical regions. A 2019 global review concluded that four main respiratory viruses had distinct seasonal patterns, and hypothesised that temperature, humidity, precipitation, and solar radiation were possible seasonal stimuli,[12] which is consistent with our findings. In our study, we found that RSV and influenza virus had clear seasonal epidemics in the winter months in northern China and biannual epidemics in southern China. Human parainfluenza virus and human bocavirus epidemics were found mostly in spring and early summer. Human metapneumovirus had an annual peak in the winter–spring period, especially in the north. Adenovirus and human coronavirus did not exhibit clear annual seasonality. A study in the USA also found that RSV had strong seasonality with annual epidemics in winter, and vapour pressure had the highest explanatory power for this seasonality.[31] A study in Sweden[20] (temperate climate) found that RSV, influenza virus, human coronavirus, and human metapneumovirus all had strong seasonal patterns in winter. A Malaysian study also found clear seasonality for common respiratory viruses.[32]

Our study found that the most dominant factor for any respiratory virus was vapour pressure, explaining up to 64% of viral infections in the north and 50% in the south. A possible explanation for the high influence of this factor is that low vapour pressure can increase the possibility of airborne virus spread by favouring rapid evaporation of the infectious droplets.[16]

Other factors such as temperature, rainfall, atmospheric pressure, and hours of sunlight also played important roles in both temperate (north) and subtropical (south) regions in China. Relative humidity had significant explanatory power in the north but not in the south.

Various theories behind the association between meteorological factors and respiratory viruses have been proposed.[17,33] Together, the mechanisms of how meteorological factors act on respiratory viruses include four main aspects: the virus' survival, infection, and transmission, and the human immune response (figure 2). First, meteorological factors can change viral stability, viability, activity, pathogenesis, or virulence, and prolong or shorten a virus' survival time.[34] Laboratory studies have concluded that most viruses tend to survive longer in lower temperatures and lower relative humidity.[35] Some aerosol viability experiments have found that adenovirus was more stable at high relative humidity; by contrast, human parainfluenza virus was more stable at low relative humidity, whereas RSV had bimodal peak stability at 20% or 40–60% relative humidity and instability at 30%. Influenza virus was also generally more stable at lower relative humidity.[35,36] Second, infections are more likely to happen in certain weather conditions. For example, in cold weather, the decrease in temperature lowers the physical protection barrier of the human body by decreasing the secretion function of respiratory mucosa membrane and weakens the body's immune function by decreasing the erythrocyte sedimentation rate in blood, which can lead to infection. Third, weather conditions can affect the transmission routes of different respiratory viruses, such as changing human behaviour (eg, indoor crowding in poor weather and promoting circulation by going outside when the weather becomes warmer and drier),[37] and changing respiratory droplet size and deposition, airflow, and recirculation,[27] making the effects of weather conditions difficult to investigate experimentally.[33,35] After being infected by a respiratory virus, the human immune system responds by changing the host's respiratory resistance and respiratory secretion production and viscosity, among other things.[38] Lastly, weather can alter host defence mechanisms, such as cooling of the nasal airway, stimulating vitamin D metabolism, impairing mucociliary clearance, and reducing the phagocytic activity of leucocytes with inhalation of cold, dry air.

Our findings on the associations between meteorological factors and each virus could be explained by these aspects. RSV is transmitted by large particle aerosols or by direct or indirect contact in human secretions.[33,35] In winter, cold weather in temperate climates (ie, northern China) with low temperature and relative humidity would facilitate its survival and transmission. Experimental studies have shown that RSV inactivation requires a longer time when the temperature decreased.[16] Although low relative humidity has been identified as a contributing factor to the transmission and survival of the influenza virus,[39,40] no significant explanation of relative humidity was found in our study ($p>0.05$). The transmission of influenza virus via airborne routes in northern China, as in most temperate regions, might be affected by ambient humidity, which is expressed by temperature and atmospheric pressure, affecting the virus' stability and respiratory droplet size.[35] In southern China, fewer hours of sunlight reduce levels of ultraviolet B radiation, which could enhance influenza virus survival, because solar radiation is well known to be an effective sterilising agent for most infectious agents.[14] Also, reduced sun exposure could lead to vitamin D levels dropping, which might down-regulate the expression of antimicrobial peptides, and further reduce immune system functionality and increase susceptibility to seasonal viruses.[41]

Our findings suggest that human parainfluenza virus was less dependent on weather in the north than in the south, where all meteorological factors had significant explanatory power other than relative humidity and wind speed. An analysis over 11 years in Suzhou, a subtropical city of China, suggested that temperature had a positive correlation with human parainfluenza virus activity,[25] while another study found a weaker correlation between human parainfluenza virus and vapour pressure.[20] Several studies in Europe, Africa, and Asia have found rainfall to be positively correlated with adenovirus activity;[25,42] comparable results were obtained in our study for southern China. Our finding that vapour pressure was the dominant factor for human metapneumovirus in both northern and southern China agrees with a Swedish study that human metapneumovirus was associated with low temperature and vapour pressure,[20] although a French study found that low temperature and wind speed were the main drivers of human metapneumovirus seasonality in temperate regions.[26] Our results for human coronavirus were also in agreement with the Swedish study,[20] with wind speed playing an important role. Two possible explanations for this role are that strong wind might be able to accelerate the spread of human coronavirus by increasing air flow, and strong wind can cool body surface temperature, which would cause vasoconstriction in the respiratory tract mucosa and suppression of immune responses, leading to increased susceptibility to infections.

When considering the interaction of different meteorological factors on respiratory viruses, we generally found that factors with the strongest interactive effect varied by virus and region. This is because these viruses display a great deal of variety not only in their virion structure and genome composition, but also in their modes of transmission among humans. The enhanced interaction between any two variables implied that the seasonal prevalence of these viruses was not determined by a single factor. Relative humidity was the most dominant interaction factor in the north for most viruses. In the south, wind speed also interacted significantly, although it was rarely significant as an individual factor in that region. In general, meteorological factors are not independent of each other, thus the epidemic cycles of respiratory viruses could be the result of the combined environmental parameters rather than the action of any single factor.

This Article, like most previous studies, relied on outdoor meteorological variables, whereas many people might spend more time indoors than outdoors.[43] It is therefore likely that the viability and transmission of viruses should not be assessed for outdoor environmental conditions, when main transmission might happen indoors.[44] As indoor data are not widely available, outdoor climate factors were used as surrogate markers of indoor climates in this study. In this sense, meteorological factors are more linked to changes in human behaviour and living environment than direct meteorological exposures, thus indirectly influencing the spread of viruses. In temperate regions, indoor crowding has frequently been speculated as a key reason for peaks in influenza virus and RSV in cold seasons.[45,46] In subtropical regions, hot and humid air in summer means people stay in crowded, air-conditioned environments, which favour the easy transmission of viruses.[47] Notably, use of air conditioning is associated with income and wealth; there is a need to further study the relationship between economic conditions and viruses.

There are some important limitations to consider. First, the spatial representation of sentinel hospitals in this study was conducted by a national surveillance protocol.[18]

Specimens were tested in different laboratories, and the time and amount of specimen collection and the testing capacity varied by hospital, which might lead to some statistical differences. Second, our present dataset was not sufficient to support the study at the city or hospital level because specimens were collected from the first two or five patients with ALRI, either weekly or monthly, in each sentinel hospital, which limits the precision of data on the local situation at the time or each patient's individual exposures. Third, virus subtyping was not done systematically and thus subtype data were not included in our analysis, which might provide a more comprehensive picture in various regions.[48] Fourth, we only superficially explored the relationship between meteorological factors and the seasonal characteristics of these viruses; our findings were just statistically extrapolated, but the true mechanism of influence of meteorological factors on these viruses might be very complex. Finally, the association between meteorological factors and these viruses could be influenced by many confounding factors. As people prefer to stay indoors during cold or rainy days, indoor climate—particularly humidity—has been suggested as an important factor in respiratory virus transmission; future studies should consider both the role of the outdoor and indoor climate on respiratory virus infections.

In conclusion, this study systematically and quantitatively described the explanatory power of different meteorological variables on seven common respiratory viruses in both temperate and subtropical climate regions of China. We detected spatiotemporal heterogeneity for most viruses. The dominant meteorological factor varied by virus, age group, and region. More understanding of the meteorological effects on the seasonality of these viruses would be helpful in guiding government planning, such as through public health interventions, infection control practice, timing of passive immunoprophylaxis, and facilitating the development of future vaccine strategies.

**Contributors**
JW and WY contributed to the conception and design of the study. BX and JY contributed to the literature review. SL, ZL, and LW contributed to the ALRI data collection and data quality control. BX, JY, and CX collected the meteorological data. BX, JY, YL, MH, and CX cleaned, analysed, and visualised the data. JW and WY supervised the analysis and generation of results, and directly accessed and verified the data. BX and JW drafted and finalised the paper. All authors contributed to data interpretation, and reviewed and approved the final version manuscript. All authors had full access to all the data in the study and had final responsibility for the decision to submit for publication.

**Declaration of interests**
We declare no competing interests.

**Data sharing**
The meteorological data used in this study can be downloaded from http://data.cma.gov.cn. Virus data can be requested from the Chinese Center for Disease Control and Prevention. The model code can be requested from the corresponding author or downloaded from http://www.geodetector.cn.

**Acknowledgments**
This study was funded by the National Natural Science Foundation of China (grants 41531179, 42071375, and 41421001), the Ministry of Science and Technology of China (grant 2016YFC1302504), and the Technology Major Project of China (grants 2018ZX10713001-001, 2018ZX10713001-011). SL acknowledge supports from the National Natural Science Fund (81773498), the National Science and Technology Major Project of China (2016ZX10004222-009), and the Program of Shanghai Academic/Technology Research Leader (18XD1400300).

Editorial note: the *Lancet* Group takes a neutral position with respect to territorial claims in published maps and institutional affiliations.